\documentclass[aps,pra,showpacs,twocolumn,superscriptaddress]{revtex4}
\usepackage{amssymb,amsthm,amsmath}
\usepackage{epsfig}
\usepackage{graphicx,color}
\usepackage{amsmath}
\usepackage{array}
\usepackage{multirow}
\usepackage{booktabs}
\usepackage{threeparttable}
\usepackage{slashed}

\begin{document}
\title{An oscillating Casimir potential between two impurities in a spin-orbit coupled Bose-Einstein condensate}
\author{Pei-Song He}
\affiliation{School of Science, Beijing Technology and Business University, Beijing 100048, China}
\author{Qing Sun}
\email{sunqing@cnu.edu.cn}
\affiliation{Department of Physics, Capital Normal University, Beijing 100048, China}
\author{An-Chun Ji}
\affiliation{Department of Physics, Capital Normal University, Beijing 100048, China}
\date{\today}
\begin{abstract}

We study the Casimir potential between two impurities immersed in a spin-orbit coupled Bose-Einstein condensate (BEC) with plane-wave order. We find that, by exchanging the virtual phonons/excitations, a remarkable anisotropic oscillating potential with both positive and negative parts can be induced between the impurities, with the period of the oscillation depending on the spin-orbit coupling strength. As a consequence, this would inevitably lead to a non-central Casimir force, which can be tuned by varying the strength of spin-orbit coupling . These results are elucidated for BECs with one-dimensional Raman-induced and two-dimensional Rashba-type SOC.

\end{abstract}
\maketitle


\section{Introduction}

The research on quantum impurity problem in a many-body system provides a promising way to study the few-body and many-body physics in condensed matter physics \cite{WilsonRMP1975, NaidonReview, DengRMP2010, NarozhnyRMP2016, AlloulRMP2009, AbrahamsRMP2001, LeeRMP2006, BalatskyRMP2006, EversRMP2008, NagaosaRMP2010}. The interplay between the impurity and the background many-body ground state can lead to many intriguing phenomena \cite{HuPRL2013, CovaciPRL2009, ShchadilovaPRL2016, JorgensenPRL2016, ShadkhooPRL2015, CasalsPRL2016, HuPRL2016, SchmidtPRX2016, SauPRB2013, WrayNatPhys2011, NetoPRL2009}, such as the orthogonality catastrophe of a time-dependent impurity in ultracold fermions \cite{AndersonPRL1967, KnapPRX2012}, an electron dressing Bose-Einstein condensate by a Rydberg-type impurity \cite{BalewskiNature2013}, and Yu-Shiba-Rusinov state \cite{Yu1965, Shiba1968, Rusinov1969, GopalakrishnanPRL2015, BrydonPRB2015}, etc.
 One of the remarkable effect is that an effective interaction may be induced between impurities via exchanging the virtual excitations of the underlying ground state fluctuations, which is also known as the Casimir effect \cite{BijlsmaMJPRA2000, PAMartin2006, XLYu2009, YNishida2009, MNapiorkowski2011, JCJaskula2012}.


Recent years, with the realization of artificial gauge fields in ultracold atoms, the spin-orbit coupled quantum gases have attracted attentive studies and many interesting physics brought by SOC have been explored \cite{GoldmanN2014, ZhaiH2015}. For example, a plane-wave (PW) phase and a stripe phase can be identified in BECs with SOC \cite{LinYJ2011}. More recently, it was found that a point-like impurity moving in a BEC with Rashba SOC would experience a drag force with a nonzero transverse component \cite{HePS2014} and a Fulde-Ferrell-Larkin-Ovchinnikov-like molecular state \cite{YiPRL2012, FFLO} may exist in spin-orbit coupled Fermi gas. Based on these developments of SOC and impurity physics in ultracold atoms, a natural and important question then raises: how the Casimir potential between impurities in a BEC is affected by SOC?

To address this problem, in this paper, we calculate the instantaneous Casimir potential between two impurities immersed in a BEC with SOC. We find that, by exchanging the virtue phonons/excitations of the PW phase, the inter-impurity potential exhibits a remarkable oscillating behavior with both repulsive and attractive components. Specifically, for the Raman-induced 1D SOC, such oscillation only exists along the direction of the SOC. While for the 2D Rashba SOC, the potential oscillates in both directions with different oscillation periods: the period in the direction of the PW wave-vector is about half of the one along the perpendicular direction. Such anisotropy of the potential would inevitably lead to a non-central Casimir force between the impurities.


This paper is organized as follows: First in Sec. II, we model the system and present 
the general formalism. Then, we analyze the results and underlying physics in Sec. III. Finally in Sec. IV, we give a brief summary. 


 


\section{Model and Formalism}

\subsection{Model}

We consider two impurity atoms immersed in a two-component interacting Bose gas with SOC.  The Hamiltonian of the system can be written as
\begin{eqnarray}
  	H &=& \int d\mathbf{r} \hat{\phi}^{\dagger}(\mathbf{r}) \bigg[-\frac{\hbar^2\nabla^2}{2m_B} - \mu
 -i \sum_{i, j = x, y, z}v_{ij}\partial_{i}\sigma_j  \nonumber\\
&& + \sum_{i = x,y,z}\Lambda_i\sigma_i\bigg]\hat{\phi}(\mathbf{r}) 
+ \frac{1}{2}\int d\mathbf{r}\big[g_{\uparrow\uparrow}\hat{n}_{\uparrow}^2 + 2g_{\uparrow\downarrow}\hat{n}_{\uparrow}\hat{n}_{\downarrow} \nonumber\\
&& + g_{\downarrow\downarrow}\hat{n}_{\downarrow}^2\big] + \int d\mathbf{r} \hat{\psi}^{\dagger}(\mathbf{r}) \left(-\frac{\hbar^2\nabla^2}{2m_I} \right) \hat{\psi}(\mathbf{r}) \nonumber\\
&& 
+ g_{I}\int d\mathbf{r} \hat{n}(\mathbf{r})\hat{\psi}^{\dagger}(\mathbf{r})\hat{\psi}(\mathbf{r}).
\label{eqn_Hamiltonian}
\end{eqnarray}
Here, $\hat{\phi}(\mathbf{r})=(\hat{\phi}_{\uparrow}(\mathbf{r}), \hat{\phi}_{\downarrow}(\mathbf{r}))^{T}$ and $\hat{\psi}(\mathbf{r})$ are the annihilation operators of the two-component boson and impurity atom fields (boson or fermion) at position $\mathbf{r}$, with the mass $m_B$ and $m_I$ respectively. $\hat{n} = \hat{n}_{\uparrow} + \hat{n}_{\downarrow}$ denotes the density operator with $\hat{n}_{\uparrow} = \hat{\phi}^{\dagger}_{\uparrow}\hat{\phi}_{\uparrow}$ and $\hat{n}_{\downarrow} = \hat{\phi}^{\dagger}_{\downarrow}\hat{\phi}_{\downarrow}$. $\sigma_x$, $\sigma_y$ and $\sigma_z$ are the Pauli matrices. $\mu$ is the chemical potential of bosons. $v_{ij}$ and $\Lambda_i$ ($i, j=x, y, z$) describe the strength of effective SOC and magnetic field along $i$-direction.  
 $g_{\uparrow\uparrow}$ ($=g_{\downarrow\downarrow}$) and $g_{\uparrow\downarrow}$ are the intra- and inter-component interactions between bosons, while the impurity atoms couple to the bosons via a density-density interaction with strength $g_I$. In this paper, we are interested in the unique features of the induced interaction between impurities introduced by the SOC, which is shown to be independent with the statistics of the impurity atom. Further for simplicity, we have neglected the possible direct interaction between impurity atoms, which would not affect the main physics essentially.  

\subsection{Casimir potential between two impurities}

Before proceeding, let us briefly discuss the effects brought by SOC. As well known that in the absence of impurities, the presence of SOC would largely change the single-particle spectrum, and give rise to degenerate single-particle ground states, e.g. double minima for Raman-induced 1D SOC and ring degeneracy for Rashba SOC. Here, the 1D SOC (along $x$-direction) corresponds to $v_{xz} = -\hbar^2 k_L/m_B$, $\Lambda_x = \Omega$ and the Rashba SOC has $v_{xx} = v_{yy} = \hbar^2 k_L/m_B$ with $k_L$ and $\Omega$ the strength of SOC and Raman coupling respectively. the SOC strength.  In both cases, all of the other $v_{ij}$ and $\Lambda_i$  $(i, j = x, y, z)$ are zero.  As a result, the many-body ground state for a homogeneous Bose-Einstein condenstate could be in a PW phase or a stripe phase, depending on the atomic interaction parameter $\eta=g_{\uparrow\downarrow}/g_{\uparrow\uparrow}$ \cite{LinYJ2011, WZheng2012, CWang2010}. Correspondingly, the excitation spectrum of each phase would be also changed dramatically. When the impurities get involved, they would interplay with such excitations of the background condensate, and a unique SOC-dependent interaction may be induced between the impurities. To show this more concretely, we take the PW ground state as an example in the following.

In the path-integral formalism, we write the partition function of Hamiltonian (\ref{eqn_Hamiltonian}) as 
$\mathcal{Z} = \int \mathcal{D}[\bar{\phi}, \phi, \bar{\psi}, \psi] e^{-\frac{1}{\hbar}S[\bar{\phi}, \phi, \bar{\psi}, \psi]}$, with $S[\bar{\phi}, \phi, \bar{\psi}, \psi] = \int^{\beta}_0 d\tau \left[\hbar\bar{\phi}\partial_{\tau}\phi + \hbar\bar{\psi}\partial_{\tau}\psi + H(\bar{\phi}, \phi, \bar{\psi}, \psi) \right] $. For a low average density of the impurity atoms and moderate impurity-boson interactions, it is assumed that we can safely neglect the modifications on the properties of the condensate as well as the dispersion of the excitations due to the impurities \cite{BijlsmaMJPRA2000}.
In this way, we further write the boson fields as $\mathbf{\phi}(\mathbf{r},\tau) = \mathbf{\phi}_0(\mathbf{r}) + \delta\mathbf{\phi}(\mathbf{r},\tau)$, where $\mathbf{\phi}_0(\mathbf{r})$ is the wave function of the condensate and $\delta\mathbf{\phi}(\mathbf{r},\tau)$ are quantum fluctuations above the condensate. In the PW phase, we have $\mathbf{\phi}_0(\mathbf{r}) = \sqrt{n_0}(u, v)^T e^{i\mathbf{k}_0\cdot\mathbf{r}}$ with $\mathbf{k}_0$ the condensed momentum of the plane wave. $n_0$ is the density of the condensed bosons and $u$, $v$ are the relative amplitudes of each component satisfying $|u|^2+|v|^2=1$.  

To facilitate the following discussions, we turn to the momentum-frequency space via the Fourier transformation
\begin{eqnarray}
\delta\phi(\mathbf{r},\tau)&=& (\hbar\beta V)^{-1/2} e^{i\mathbf{k}_0\cdot\mathbf{r}}\sum_{\mathbf{q},\nu}\delta\phi(\mathbf{q},\nu) e^{i\mathbf{q}\cdot\mathbf{r}-i\nu\tau},\\
\psi(\mathbf{r},\tau)&=&(\hbar\beta V)^{-1/2}\sum_{\mathbf{k},\omega}\psi(\mathbf{k},\omega) e^{i\mathbf{k}\cdot\mathbf{r}-i\omega\tau}.
\end{eqnarray}
Here, $\mathbf{q}$ (relative to the condensed momentum $\mathbf{k}_0$) and $\mathbf{k}$ are the momenta of the quantum fluctuation $\delta\phi(\mathbf{q},\nu)$ and the impurity field $\psi(\mathbf{k},\omega)$, respectively. $\nu$ and $\omega$ are Matsubara frequencies in the imaginary time.


Within the Bogoliubov approximation of a weakly interacting gas, the partition function can be expanded up to second order, which becomes $\mathcal{Z} = \int \mathcal{D}[\delta\bar{\phi}, \delta\phi, \delta\bar{\psi}, \delta\psi] e^{- \frac{1}{\hbar}[S_0 + S^{(2)}]}$. Here, $S_0$ is the action in the classical level, and $S^{(2)}$ is the Gaussian part, which is given by 
\begin{eqnarray}
 	S^{(2)} &=& \frac{1}{2}\sum_{\mathbf{q},\nu}\delta\bar{\Phi}(\mathbf{q},\nu)\mathbf{M}(\mathbf{q},\nu)\delta\Phi(\mathbf{q},\nu)\nonumber\\
 	&& + \sum_{\mathbf{k},\omega}\frac{\hbar^2\mathbf{k}^2}{2m_I}\bar{\psi} (\mathbf{k},\omega)\psi (\mathbf{k},\omega)\nonumber\\
 	&&+\frac{1}{2} \sum_{\mathbf{q},\nu}\big[\mathbf{J} ^{\dagger}(\mathbf{q},\nu)\delta\Phi(\mathbf{q},\nu)+h.c.\big],
\label{Z_Gaussian}
\end{eqnarray}
where $\delta\Phi(\mathbf{q},\nu) = \big[\delta\phi_{\uparrow}(\mathbf{q},\nu),\delta\phi_{\downarrow}(\mathbf{q},\nu),\delta\bar{\phi}_{\uparrow}(-\mathbf{q},-\nu)$, $\delta\bar{\phi}_{\downarrow}(-\mathbf{q},-\nu)\big]^T$, $\mathbf{J}(\mathbf{q},\nu) = g_I\sqrt{n_0}\rho(\mathbf{q},\nu)(u,v,u,v)^T$ with $\rho(\mathbf{q},\nu) = \sum_{\mathbf{k},\omega}\bar{\psi}(\mathbf{k},\omega)\psi(\mathbf{k}+\mathbf{q},\omega+\nu)$, and
\begin{equation}
	\mathbf{M}(\mathbf{q},\nu) 
	= -i\hbar\nu\begin{pmatrix}I&0\\0&-I\end{pmatrix} + \begin{pmatrix}A_{\mathbf{q}}+B&B\\B&A^{\dagger}_{-\mathbf{q}}+B\end{pmatrix}
\end{equation}
with $I$ the $2\times2$ unit matrix and 
\begin{eqnarray}
A_{\mathbf{q}}&=&H_0(\mathbf{q} + \mathbf{k}_0)I  - \mu I +\eta gn_0 I \nonumber\\
&&-(\eta-1)gn_0 \begin{bmatrix}u^2&0\\ 0&v^2\end{bmatrix},\nonumber\\
B&=&gn_0\begin{bmatrix}u^2&\eta uv\\\eta uv&v^2\end{bmatrix}.
\end{eqnarray}
Here, $H_0(\mathbf{q}) = \frac{\hbar^2\mathbf{q}^2}{2m_B} + \sum_{i, j = x, y, z}v_{ij}q_i\sigma_j   + \sum_{i = x,y,z}\Lambda_i\sigma_i$.

We further introduce the time-ordered Green's function of the boson fields $\delta{\Phi}$ as 
\begin{equation}
 	\mathcal{G}_0(\mathbf{r},\tau;\mathbf{r}',\tau') \equiv -\frac{1}{2\hbar}\langle 0| T_{\tau}\delta\Phi(\mathbf{r},\tau)\delta\bar{\Phi}(\mathbf{r}',\tau')|0\rangle,
\end{equation}
and its Fourier transformation 
\begin{eqnarray}
 	\mathcal{G}_0(\mathbf{r},\tau;\mathbf{r}',\tau') &=& \frac{1}{\hbar\beta V}\sum_{\mathbf{q},\nu}\mathcal{G}_0(\mathbf{q},\nu)e^{i\mathbf{q}\cdot(\mathbf{r}-\mathbf{r}')-i\nu(\tau-\tau')}, \nonumber\\
	\mathcal{G}_0(\mathbf{q},\nu)  &=& \left[-\mathbf{M}(\mathbf{q},\nu)\right]^{-1}.
\end{eqnarray}
Here, $|0\rangle$ is the vacuum state in the quasi-particle basis. Then the effective action for impurities can be obtained by integrating out the fluctuations of boson fields in Eq. (\ref{Z_Gaussian}), which gives 
\begin{eqnarray}
 	S_{\rm eff} &=&  \int^{\hbar\beta}_{0}d\tau \int d\mathbf{r} \bar{\psi}(\mathbf{r}, \tau) \left(\hbar\partial_{\tau}-\frac{\hbar^2\nabla^2}{2m_I} \right) \psi(\mathbf{r}, \tau) \nonumber\\
&&+ \frac{1}{2} \int^{\hbar\beta}_{0}d\tau d\tau' \int d\mathbf{r} d\mathbf{r}'\nonumber\\
 	&&\times |\psi (\mathbf{r},\tau)|^2 V (\mathbf{r},\tau;\mathbf{r}',\tau') |\psi (\mathbf{r}',\tau')|^2,
\label{eqn_Cas_1}
\end{eqnarray}
where
\begin{eqnarray}
 	V (\mathbf{r},\tau;\mathbf{r}',\tau') &=& \frac{1}{\hbar\beta V}\sum_{\mathbf{q},\nu}V (\mathbf{q},\nu) e^{i\mathbf{q}\cdot(\mathbf{r}-\mathbf{r}')-i\nu(\tau-\tau')},\nonumber\\
 	V (\mathbf{q},\nu) 
 	&=& g_I^2 n_0 (u,v,u,v)\mathcal{G}_0(\mathbf{q},\nu)(u,v,u,v)^T .\nonumber\\
\label{eqn_V_q_nu}
\end{eqnarray}




Above, Eq. (\ref{eqn_V_q_nu}) is the main start-point of this paper. In general, the fluctuations of the condensate can mediate an interaction between impurities. And we are interested in the instantaneous component of $\nu=0$ \cite{BijlsmaMJPRA2000}, which is justified by the dominant processes of emitting and absorbing of virtual quasi-particles between two impurities.  In this case, the Casimir potential takes the form of $V (\mathbf{r}, \tau; \mathbf{r}', \tau') \simeq V (|\mathbf{r}-\mathbf{r}'|)\delta(\tau-\tau')$, reflecting the ``vacuum" energy modification from these exchanging of quasi-particles.


To explore the main features of the induced potential, in the following, we take the Raman-induced 1D and Rashba-type SOCs as examples to demonstrate the underlying physics.

\begin{figure}[b]
\center
\includegraphics[width=0.5\textwidth]{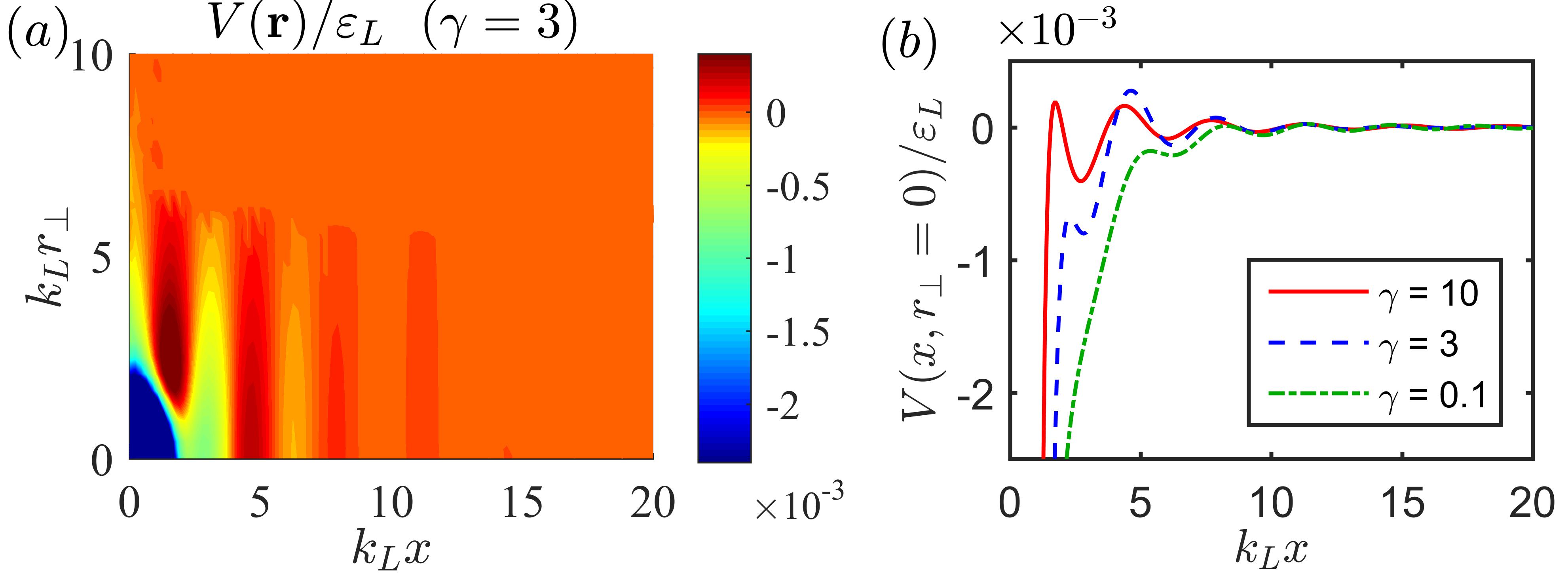}
\caption{(a). Distribution of the Casimir potential $V(\mathbf{r})$ (in unit of $\varepsilon_L$) in $x, r_\perp$-plane of the relative coordinate $\mathbf{r}$ between two impurities in a three-dimensional BEC with Raman-induced one-dimensional (along $x$-direction) SOC.  (b). The Casimir potential $V(x,r_\perp=0)$ along $x$-axis for $\gamma=10$ (red solid), 3 (blue dashed), and 0.1 (green dash-dotted). Other parameters are $\alpha=0.25$ and $\eta=1.0$.}
\label{D3_Vr}
\end{figure}

\section{Results}

\subsection{Raman-induced 1D SOC}

In this case, the PW phase survives in the parameter regime with $0<\alpha<1$ ($\alpha\equiv  \Omega m_B/\hbar^2 k_L^2$) and $\eta >\eta_c= (2-3\alpha^2)/(2-\alpha^2)$, which would transit to a zero-momentum phase at $\alpha = 1$ \cite{WZheng2012}. For $\alpha\ll 1$, we have 
\begin{eqnarray}
\left\{
\begin{array}{l}
u=-\sin\frac{\gamma_{\mathbf{k}_0}}{2}\\
v=\cos\frac{\gamma_{\mathbf{k}_0}}{2}\\
\mathbf{k}_0=-\sqrt{1-\alpha^2}k_L\mathbf{e}_x,
\end{array}
\right.
\end{eqnarray} with $\sin\gamma_{\mathbf{k}_0} = \Omega/\sqrt{(k_L k_0/m_B)^2 + \Omega^2}$.  Correspondingly, the chemical potential $\mu$ is given by 
\begin{eqnarray}
 	\mu &=& \varepsilon_{\mathbf{k}_0-\mathbf{k}_L}\sin^2\frac{\gamma_{\mathbf{k}_0}}{2} +  \varepsilon_{\mathbf{k}_0+\mathbf{k}_L}\cos^2\frac{\gamma_{\mathbf{k}_0}}{2}   - \Omega \sin\gamma_{\mathbf{k}_0}  \nonumber\\
 	&& + gn_0 +\frac{1}{2}(\eta-1)g n_0 \sin^2\gamma_{\mathbf{k}_0} - \varepsilon_L,
\end{eqnarray} 
with $\varepsilon_L \equiv\hbar^2k_L^2/2m_B$.
With Eqs. (11, 12), one can derive the explicit expression of the instantaneous Casimir potential $V(\mathbf{r})$ given by Eq. (\ref{eqn_V_q_nu}) (not shown here). In Fig. (\ref{D3_Vr}a), we plot $V(\mathbf{r})$ in the plane of the relative coordinate $\mathbf{r}$ with $\gamma=3$.  Here, $\gamma \equiv \xi^{-2}/k_L^2$ denotes the ratio between boson-boson interaction energy and the kinetic energy characterized by the strength of SOC, where $\xi= \hbar/\sqrt{2gn_0 m_B}$ is the healing length of the boson system. We find that, in contrast to that without SOC, the instantaneous Casimir potential exhibits a remarkable oscillation between repulsive and attractive parts with the varying of the distance between impurities. Furthermore, such oscillation is found to be along the $x$ direction, with period about $\pi/|\mathbf{k}_0|$, as shown in Fig. (\ref{D3_Vr}b).

\begin{figure}[t]
\center
\includegraphics[width=0.49\textwidth]{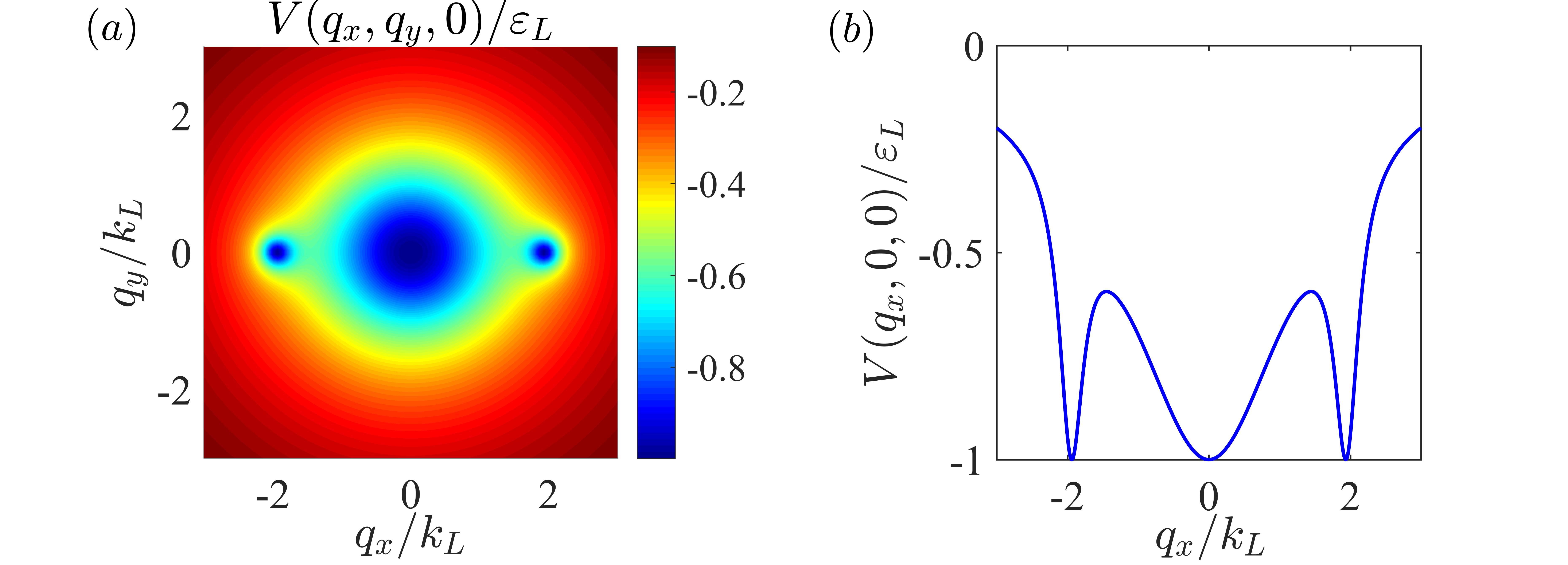}
\caption{The Fourier distribution $V(\mathbf{q})$ of the Casimir potential $V(\mathbf{r})$ given in Fig. (\ref{D3_Vr}): (a). $V(q_x, q_y, 0)$ in $q_x, q_y$-plane with $q_z=0$; (b). $V(q_x, 0, 0)$ along $q_x$-axis with $q_y=q_z=0$. Here, $\alpha=0.25$, $\eta=1$, and $\gamma=3$.  }
\label{D3_Vk}
\end{figure}

The oscillating behavior can be understood as follows. 
In the basis of single particle states of the free boson gas, the inter-impurity Casimir potential is a consequence of the scattering between the condensed bosons and the excited ones, as can be seen in the last term of Eq. (\ref{Z_Gaussian}). In the static limit of impurity potential (that is, $\nu = 0$), it prefers scattering process between states with small energy difference. To see it more clearer, in Fig. (\ref{D3_Vk}), we give the Fourier component $V(\mathbf{q})$ of the Casimir potential. 
One can see that, for the Raman-induced 1D SOC, $V(\mathbf{q})$ is domianted by the excited states around the condensed momentum and those around the other degenerate momentum, with the exchanged momenta about zero and $\pm2\mathbf{k}_0$, respectively. The distribution of $V(\mathbf{q})$ around zero momentum are common to that of the BEC without SOC, and it will contribute to an attractive, divergent, and damping Casimir potential. While those around $\pm2\mathbf{k}_0$ are the unique feature brought by the Raman-induced 1D SOC, which give rise to the oscillations of the potential with a period about $\pi/|\mathbf{k}_0|$. It is noteworthy that, this oscillating behavior bears some similarity with the well-known Ruderman-Kittel-Kasuya-Yosida (RKKY) indirect exchange between two localized spins. The main difference is that, the RKKY interaction is mediated by the delocalized fermions around the Fermi surface \cite{RKKY}. While in our case, the induced interaction is mediated by the bosonic excitations around the ground state with finite momentum. Moreover, due to the macroscopic occupancy of the condensed state, there is a bosonic enhancement $n_0$ in the Casimir potential, making the oscillations prominent. 


We also find that the range of the Casimir potential is on the order of the healing length $\xi$, similar to that without SOC. Since the period of oscillation is about $\pi/|\mathbf{k}_0|$, the number of oscillations gets lesser for smaller $|\mathbf{k}_0|$, which can be achieved by making $k_L$ smaller or $\alpha$ larger.  Furthermore, as $\xi$ gets larger, that is for larger $\gamma$, the positive humps at small relative distance gradually becomes negative humps and even vanish, as shown in Fig. \ref{D3_Vr}(b).

The amplitude of the oscillations grows quickly as the strength of SOC gets larger, as shown in Fig. \ref{D3_Vr} in which the the Casimir potential is scaled with $\varepsilon_L$.
So the Casimir potential can be prominent for large enough SOC strength.
We have also calculated the Casimir potential in 2D with Raman-induced 1D SOC.
By comparing the results in 2D ans 3D systems, we find that the amplitude of the oscillation in Casimir potential is more prominent in lower dimensional systems.  It is because that the ratio of the those scatterings contribute to the oscillations to all of the scatterings is larger in lower dimension.
We also find that as $\eta$ increases, the amplitude of the oscillations becomes diminished.  This is because the gap of excitations which corresponds to momentum transfer around $\pm2\mathbf{k}_0$ along the $k_x$ axis goes larger when $\eta$ is increased, and it results in smaller scattering probability with corresponding transfered momentum.

Another important feature of the Casimir potential is that it is anisotropic due to the anisotropic excitation spectrum, which means that the Casimir force between the two impurities is non-central.  Similar behavior has also been found previously in drag force experienced by a moving impurity in BEC with SOC \cite{HePS2014}.

\begin{figure}[t]
\center
\includegraphics[width=0.5\textwidth]{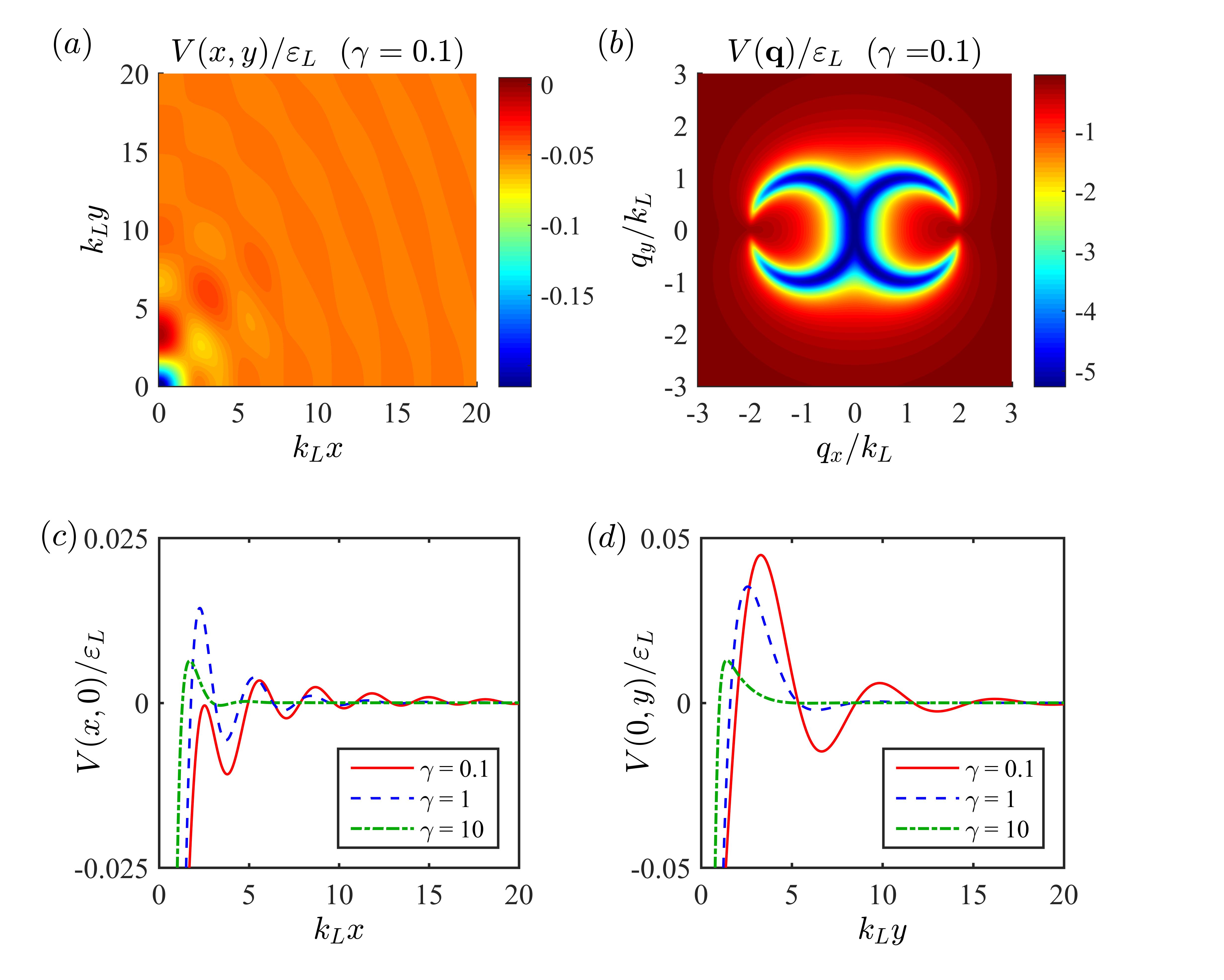}
\caption{(a) The real space distribution $V(x, y)$ and (b) the momentum space distribution $V(q_x, q_y)$ of the Casimir potential between two impurities in a two-dimensional BEC with Rashba-type SOC.  (c) and (d) are the Casimir potential along $x$- and $y$-axis respectively, with $\gamma=0.1$ (red solid), 1 (blue dashed), and 10 (green dash-dotted). $\eta = g_{\uparrow\downarrow}/g = 0.9$.}


\label{Rashba_V}
\end{figure}

\subsection{Rashba-type SOC}


For isotropic Rashba SOC \cite{CWang2010}, the PW phase appears for $\eta \le 1$. We have
\begin{eqnarray}
\left\{
\begin{array}{l}
u=v=\frac{1}{\sqrt{2}}\\
\mathbf{k}_0=-k_L\mathbf{e}_x,
\end{array}
\right.
\end{eqnarray}
and 
\begin{equation}
 	\mu = - \frac{\hbar^2 k_0^2}{2m_B} + gn_0 +2(\eta-1)gn_0 u^2  v^2.
\end{equation}
After some straightforward calculations, the dimensionless Casimir potential takes
\begin{equation}
 	V (\mathbf{r}) = \int\frac{d^2\mathbf{q}}{(2\pi)^2}V (\mathbf{q}) \cos(\mathbf{q}\cdot \mathbf{r}),
\end{equation}
where
\begin{eqnarray}
 	V (\mathbf{q}) &=& - \frac{2g_I^2}{g} \gamma\cdot\big\{q^6 - 4  q^2(2q_x^2 - q_y^2) + 16 q_x^2 \nonumber\\
 	&& +(1-\eta) \gamma(q^4+4 q_x^2) \big\}\big/\big\{(q^4-4 q_x^2)^2 \nonumber\\
 	&& +  2\gamma\left[q_x^2(q^2-4 )^2+q^2q_y^2(q^2+4 )\right] \nonumber\\
	  &&+(1-\eta^2)\gamma^2(q^4+4 q_x^2) \nonumber\\
	  && - 4(1-\eta) \gamma\left[q^4 + 4q_x^2 (1 - q^2)\right]\big\}.
\label{eqn_Vr}
\end{eqnarray}
It is interesting to see that, above $V(\mathbf{q})$ is just the dynamical structure factor at zero Matsubara frequency \cite{HePS2012, Mahan}.



In Fig. (\ref{Rashba_V}), we plot the distributions of $V(\mathbf{r})$ and $V(\mathbf{q})$ in the real space and momentum space respectively. One can see that, the potential in this situation also shows significant oscillating behavior as in the case of Raman-induced 1D SOC. Nevertheless, there are some important differences arising from the ring degeneracy of the single-particle states brought by the isotropic Rashba SOC.

First, the Casimir potential for the Rashba SOC oscillates both along $x$- and $y$-direction as depicted in Fig. (\ref{Rashba_V}c) and (\ref{Rashba_V}d). In particular, the periods of oscillations along both directions are found to be around $\pi/k_L$ and $2\pi/k_L$, respectively. The reason is that, due to the Rashba SOC, the scattering processes with momentum transfer around $2k_L$ and $k_L$ along $x$- and $y$-direction between the impurity and the background BEC, are largely enhanced. Second, along the $y$-axis, a positive hump is developed (see Fig. \ref{Rashba_V}d).  Third, with the decrease of $\eta$, the excitation gap along $x$-direction increases. As a result, the oscillating behavior along $x$-direction gradually diminishes. While the behavior of the potential along $y$-direction is not changed.

\section{Conclusions and Discussions}

In conclusions, we have calculated the instantaneous Casimir potential between two impurities immersed in BECs with Raman-induced 1D SOC and Rashba-type SOC. We find that due to the SOC, the Casimir potential between impurities exhibits remarkable oscillations with both positive and negative components, and the period of the oscillation is inversely proportional to the strength of SOC. In addition, the amplitude of the oscillations become prominent with the increasing of the SOC strength, and can be further tuned by varying the atomic interactions. Moreover, the anisotropic potential suggests a non-central Casimir force between two impurities. Our results would be beneficial for the study of impurity physics as well as the nontrivial effects brought by SOC.



Up to now, we have only considered the single-phonon process and neglected the possible multi-phonon process to obtain the instantaneous Casimir potential. This is valid for the weakly interacting two- and three-dimensional Bose gases at zero temperature (our case), where the fraction of the quantum depletion is rather small with a much smaller possibility of the multi-phonon process. On the other hand, for 1D BEC at finite temperature, the quantum depletion becomes quite large, and the multi-phonon process can not be neglected. It is pointed out that, when the two-phonon exchanging process is included, the potential would become long-ranged \cite{SchecterMPRL2014}. We leave the SOC effect on the Casimir potential with multi-phonon process for the future study.


Experimentally, BECs with Raman-induced 1D SOC have been realized in ultracold atom experiments \cite{LinYJ2011,ZhangJY, WangPJ, Cheuk, Qu, LiJ} and many proposals on two-dimensional isotropic Rashba SOC have been proposed \cite{QSun2015, SWSu2016, DLCampbell2016}. Moreover, techniques in detecting dynamics and correlations with single atom resolution \cite{OttH2016} have also been achieved.  With these advances, our theoretical results may be verified in the near future. 

\section*{Acknowledgements}
We acknowledge the helpful discussions with H. Pu and J. M. Zhang. This work was supported by the National Natural Science Foundation of China (NSFC) under Grants Nos. 61405003, 11404225, 11474205 and Scientific Research Project of Beijing Educational Committee under Grants Nos. KM201510011002, KM201510028005.

\end{document}